\documentclass[a4paper,11pt]{article}
\usepackage{pos}
\usepackage{float}

\usepackage{lineno}

\title{Two-zone model as origin of hard gamma-rays spectrum in extreme BL Lacs}
 \ShortTitle{Two-zone model}

\author*[a]{E. Aguilar-Ruiz}
\author[a]{N. Fraija}

\affiliation[a]{Instituto de Astronom\'ia, Universidad Nacional Aut\'onoma de M\'exico, Ciudad de Mexico, Mexico}

\emailAdd{eaguilar@astro.unam.mx}
\emailAdd{nifraija@astro.unam.mx}

\abstract{The emission of the so-called extreme blazars challenges the particle acceleration models. The hardness of its spectrum, $<2$, demands extreme parameters using the standard one-zone SSC model in the high energy band. Some authors use both two-zone or hadronic/leptohadronic models to relax these extreme values. In this work, we present a leptohadronic two-zone model with external radiation fields to explain the broadband emission, where the contribution of two components forms the hard-spectrum in the $\gamma$-rays band. The first is produced by the photopion process, where accelerated protons in an inner blob located close to the core interact with the X-ray photons coming from a pair plasma. This mechanism will be responsible for $\gamma$-rays in the TeV's energies range. The second contribution is produced by an outer blob, which corresponds to the source of X-rays and $\gamma$-rays with sub-TeV energies via the standard SSC and EIC model. We exemplify our model with the prototype extreme blazar 1ES 0229 +200 obtaining a good description of its spectral energy distribution.
}

\FullConference{37$^{\rm{th}}$ International Cosmic Ray Conference (ICRC 2021)\\
		July 12th -- 23rd, 2021\\
		Online -- Berlin, Germany}

\begin{document}
\maketitle
\section{Introduction}
The so-called extreme blazars are BL Lac objects that can exhibit this property through its spectral energy distribution (SED) into two ways: in the form of extreme synchrotron with a low-bump frequency peaking above $\geq 10^{17} \, {\rm Hz}$, i.e., \textit{Extreme Synchrotron peak} (EHSP or EHBL) \citep{2001A&A...371..512C} or in the form of extreme gamma-rays with a high-bump energy peaking above $> 2-10 \rm \, TeV$, i.e., \textit{Extreme-TeV BL Lacs} \citep{Tavecchio&etal_2011} or \textit{Hard-TeV BL Lacs} (TBL) \citep{Costamante2018}. A BL Lac that is a EHBL can also have a TBL behavior at same time, but this co-existence is not a requirement.  Furthermore, observations shows that BL Lacs can show the extreme condition for time's lag suggesting that the extreme behavior is a transient stage (e.g., see \citep{Ahnen2018}). Additionally, Foffano et al. (2019)  suggested that TBL is not a homogeneous class because they found different spectral properties in the TeV-band \cite{Foffano&etal_2019}.
Modelling the SED of extreme synchrotron-peak blazars  does not represent a problem by the standard shock acceleration mechanism with one-zone emission geometry. However, there are many problems when the TeV peak (for TBL case) is modelled  with the standard one-zone synchrotron self-Compton (SSC). This model for the case of EHBL/TBL (i.e. those with EHBL and TBL simultaneous behaviour) demands parameters with unusual values (e.g., see \cite{Costamante2018, 2015MNRAS.448..910C} and references therein). 

With the goal of avoiding the extreme parameters that is demanded by SSC one-zone model, we propose in this work a lepto-hadronic two-zone model to explain the broadband SED of EHBL/TBL.

\section{Theoretical model}\label{sec:Model}	
We propose the contribution from different emission's zones  as the origin of the TBL spectrum. We focus on two main regions described by a standard blob geometry emission which we name as inner and outer blob. Moreover, as seed photons that interact with accelerated electron/proton $e^\pm$-plasma in the blob, we consider as external sources those photons that emerge from the Broad-Line Region (BLR) and the Dusty Torus (DT).  The inner blob is located very close to the SMBH at s distance $r_i$ and accelerate particles in it which interact with the radiation coming from the pair-plasma. The outer blob instead is located far away from SMBH at a distance of $r_o$ and electrons in it interact with the radiation from the BLR and the DT region. 

In order to express all quantities we define 4 references frame: the observed, the pair-plasma and the blob (inner or outer) frames. Here we denote with Latin Capital Letter with the superscript $' \rm ob '$ for observed quantities while the AGN frame will be without the superscript. The plasma, inner blob and outer blob will be denoted with Greek lowercase Letters using unprime , prime or two-prime, respectively. For example, the observed energy is $E^{\rm ob}$ and the energy measured in the comoving frame of the AGN, the pair-plasma, the inner blob and the outer blob are $E$, $\varepsilon$, $\varepsilon^\prime$, $\varepsilon^{\prime\prime}$, respectively.
Here, we consider the on-axis case for relativistic blobs, i.e., the viewing angle is $\theta_{\rm obs} \lesssim 1/\Gamma$ and the Doppler factor is $\mathcal{D} = \left[ \Gamma(1-\beta\cos{\theta_{\rm obs}})\right]^{-1} \simeq 2 \Gamma$. We assume different speed for each blob, i.e., $\Gamma_o \neq \Gamma_i$. In the following we described each zone in more detail.
\subsection{Pair-plasma}
Near the SMBH MeV-photons annihilate with lower energy photons ( i.e.  $\gamma+\gamma \rightarrow e^\pm$), producing an $e^\pm$ outflow that moves with velocity $\beta_{\rm pl} \approx 0.3 - 0.7$. These pairs produce an optically thick environment if the disc luminosity above $\rm 511 \, {\rm keV}$  is greater than $\gtrsim 3 \times 10^{-3} \, L_{\rm E}$, where $L_E \approx 1.26 \times 10^{47} \, {\rm erg/s}  \, ( M_{\bullet}/10^9 M_{\odot} )$ is the Eddington's Luminosity and $M_{\bullet}$ is the mass of the SMBH. The radiation only escapes at the photosphere's radius, $R_{\rm ph}$,  producing a line shape spectrum peaking at $\varepsilon_{\rm pl} \approx 511 \, {\rm keV}$ \citep{1999MNRAS.305..181B}. The photosphere occurs at a compact radius which is of the order of Schwarzschild radius $R_g \approx 1.5 \times 10^{14} \, {\rm cm} \, ( M_{\bullet}/10^9 M_{\odot})$. The released radiation is very anisotropic where the most significant contribution is confined within a solid angle $\Omega_{\rm pl} \lesssim 0.2 \pi$. Here, we assume the emission's shape is very narrow such that its photon distribution can be described by a delta approximation as

\begin{equation}\label{eq_line_distribution}
    n_{\rm pl} (\varepsilon) = \frac{u_{\rm pl} }{\varepsilon_{\rm pl}}\, \delta(\varepsilon - \varepsilon_{\rm pl}),
\end{equation}

where the energy density is $u_{\rm pl} = L_d / (\Omega_{\rm pl} R_{\rm ph}^2 \beta_{\rm pl } c)$.
If the BLR's luminosity represents a fraction of disc's luminosity, i.e., $L_{\rm BL} = \phi_{\rm BL} \, L_d$ therefore for BL Lacs we have an upper value given by $L_d \lesssim 5 \times 10^{-4} \phi_{\rm BL}^{-1} \, L_{\rm E}$  \citep{Ghisellini&etal_2011}.

\subsection{The inner blob}

The inner blob is a compact region with size of the order of few times $R_g$ and located very close to the plair-plasma photosphere.   
This region is moving away respect to the pair-plasma, then photons coming from the pair-plasma are observed in the blob with an energy and an energy density of 
\begin{equation}\label{eq_plasma_energy}
   \varepsilon'_{\rm pl} = \varepsilon_{\rm pl}/(2\Gamma_{\rm rel}) \, , \qquad {\rm and} \qquad
    u^{\prime}_{\rm pl}  = u_{\rm pl}/(2\Gamma_{\rm rel})^2 \, ,
\end{equation}
respectively, where $\Gamma_{\rm rel}$ is the relative Lorentz factor between the pair-plasma and the inner blob, which is given by
\begin{equation}\label{eq_GammaRel}
\Gamma_{\rm rel} = \Gamma_i \Gamma_{\rm pl} \left( 1 - \beta_i \beta_{\rm pl} \right)\,.
\end{equation}

\paragraph{TeV gamma-rays.}
The presence of accelerated protons in the inner region inevitably interact with photons coming from the pair-plasma  via photopion channel producing gamma-rays.  The main contribution comes from $\Delta^+$-resonance channel, $p+\gamma \rightarrow \Delta^+ $ decaying into $\Delta^+ \rightarrow p + \pi ^0 (n + \pi^+)$. Following, pions decay into stable particles into $\pi^0 \rightarrow 2\gamma$  and $\pi^+ \rightarrow \nu_\mu + \Bar{\nu}_\mu  + \nu_e + e^+$. The proton energy threshold to pion production in the head-on collision is ${\epsilon'}_{p,\rm th}^{p\pi} = ({ m_{\Delta}^2 - m_p^2 })/{4 \epsilon'}$ and the produced gamma-rays carries out 10\% of the proton energy. Therefore, observed gamma-rays produced with the pair-plasma photons are expected above
\begin{equation}\label{eq_TeV_gamma_rays}
{E}^{\rm obs}_{\gamma,p \pi} \gtrsim 900 \, {\rm GeV} \, (\mathcal{D}_i/10) \, (\Gamma_{\rm rel}/3) \, .
\end{equation}

\paragraph{Emerging gamma-rays.}
The gamma-ray spectrum produced in the inner region will be attenuated by photons from the pair-plasma as well as the radiation produced in the outer blob. The optical depth is given by

\begin{equation}\label{eq_optical_depth}
    \tau_{\gamma\gamma}(E_\gamma) \simeq  \frac{L}{2c}\int_{-1}^{+1} d\mu (1-\mu) \int_{\varepsilon_{th}}^\infty dE \; \sigma_{\gamma \gamma}(\beta_{\rm cm}) \; n_{\gamma}(E)\,,
\end{equation}
where $\sigma_{\gamma\gamma}$ is cross-section, $\varepsilon_{\rm th} $ is the threshold energy for pair creation and $\beta_{cm}$ is the created pair's velocity in the center of mass frame \citep{Dermer&Menon_2009}, $\mu$ the collision angle and $L$ is the mean distance that photons travel. For example, for pair-plasma photons that permeate the inner blob the distance is the inner blob size, for BLR is the width of the shell of BLR whereas for the dusty torus is the size of the torus.
\begin{figure*}
\begin{minipage}[b]{0.50\linewidth}
\centering
\includegraphics[width=0.9\linewidth]{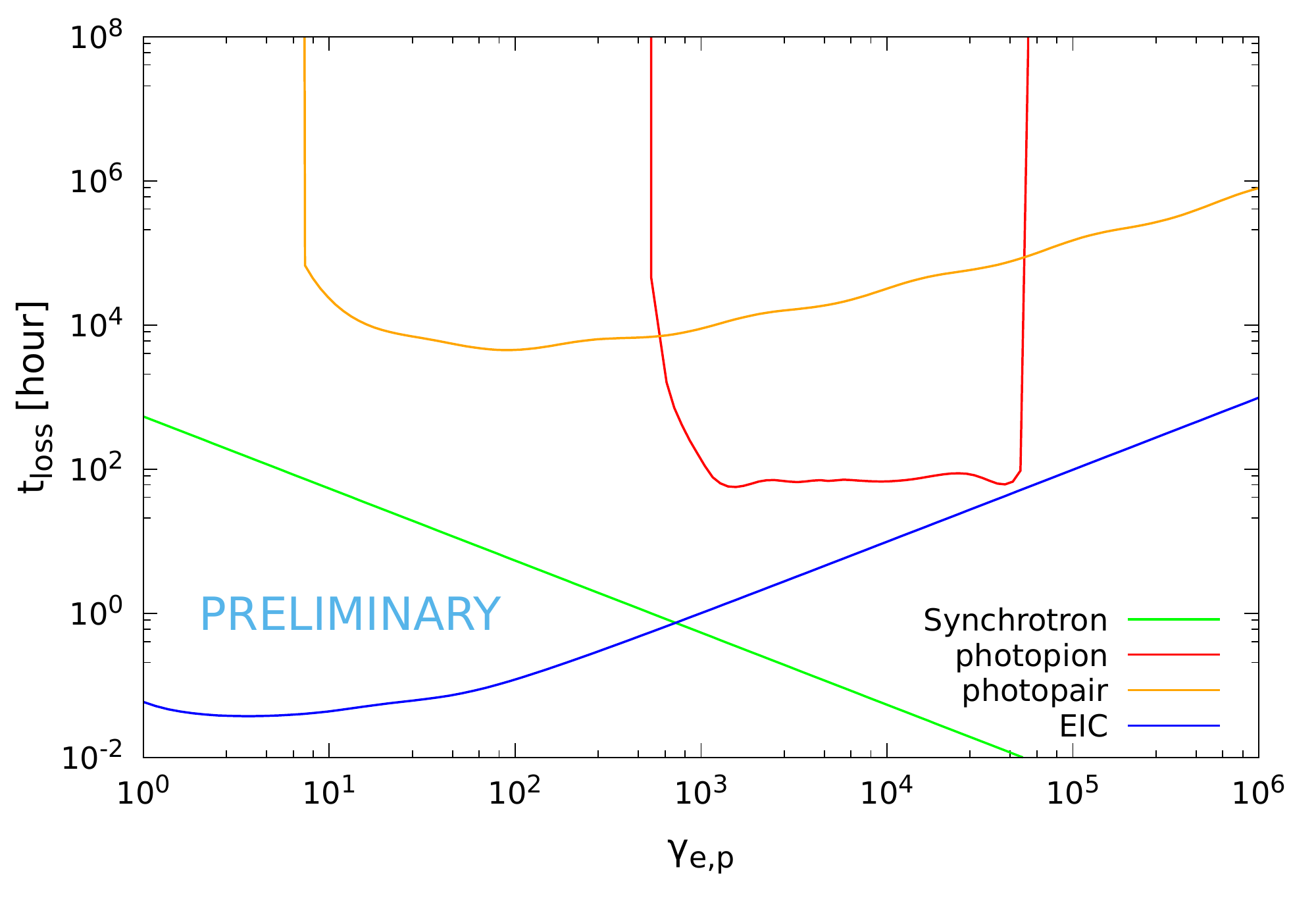}
\end{minipage}\hfill 
\begin{minipage}[b]{0.50\linewidth}
\centering
\includegraphics[width=0.90\linewidth]{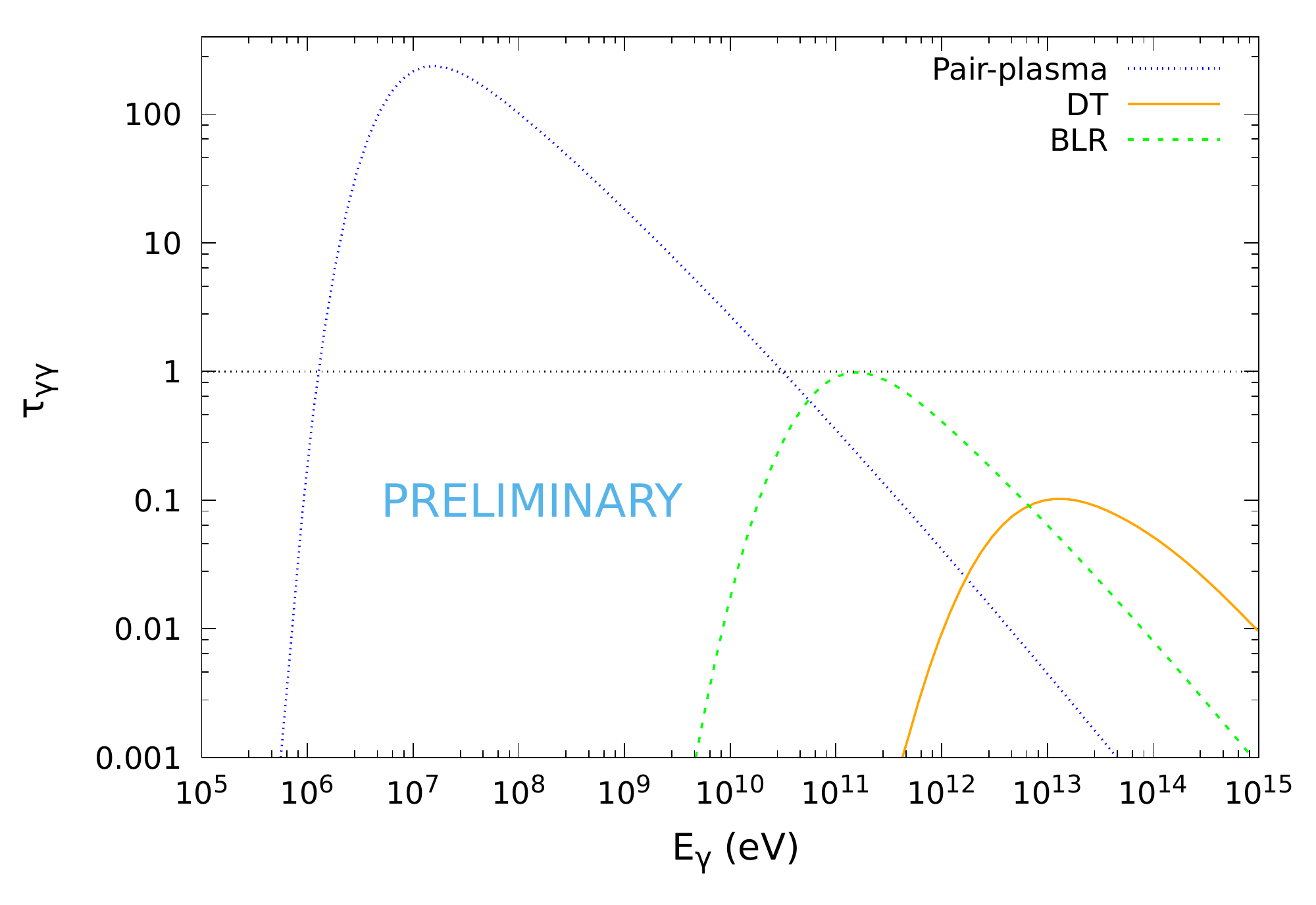}
\end{minipage} 

\caption{ \textbf{Left)} Timescales for electrons and protons calculated in the inner blob. We consider external seed photons coming from the pair-plasma assuming a SMBH mass of $M_\bullet = 10^9 M_\odot$, disc luminosity $L_d =  3\times10^{-3}L_E$, Boost Lorentz factor of $\Gamma_{i}=3$ and a magnetic field of $B_i=20 \, \rm G$. \textbf{Right)} Optical depth for $\gamma\gamma$ absorption considering different attenuation sources. In the case of the pair plasma we use the same values used for the timescales calculation.}
\label{fig_optdepth}
\end{figure*}

\subsection{Outer blob }
External radiation from the BLR and the DT plays an important role because due to the interactions with electrons in the blob. The location of BLR and DT is given by the relationship 
\begin{equation}\label{eq_rBL}
    r_{\rm BL} \approx 10^{17} \, {\rm cm} \; \left( {L_{\rm d}}/ {10^{45} \, \rm erg \, s^{-1}} \right)^{1/2} , \qquad {\rm and}
    \qquad
     r_{\rm DT} \approx 2.5 \times 10^{18} \, {\rm cm} \;  \left( {L_{\rm d}}/ {10^{45} \, \rm erg \, s^{-1}} \right)^{1/2} \, ,
\end{equation}
respectively.  Using the previous relations in the AGN scenario, the energy densities are given by \citep{Hayashida&etal_2012}

\begin{equation}\label{eq_energy_densities}
    u_{\rm BL} \approx 0.26  \, {\rm \frac{erg}{cm^{3}}} \; \frac { \phi_{\rm BL}} {1 + (r_o/r_{\rm BL})^3  }\, \qquad {\rm and}
    \qquad
    u_{\rm DT} \approx 4.3 \times 10^{-3}  \, {\rm \frac{erg}{cm^{3}}} \; \frac { \phi_{\rm DT}} { 1 + (r_o/r_{\rm DT})^4 }\, ,
\end{equation}
 where the terms of $\phi_{\rm BL}$ and $\phi_{\rm DT}$ are the fraction of disc luminosity reprocessed by the BLR and the DTs, respectively.

\section{Application: The prototype 1ES 0229+200}\label{sec:Results}

Located at a redshift of $z=0.136$, the prototype EHBL/TBL is 1ES 0229+200. This blazar harbour a SMBH with a mass of $M_{\bullet} = 10^{9.16 \pm 0.11}\,{M_{\odot}}$ . Its harder spectrum in the VHE band was discovered by HESS observatory.  We apply the model described in the previous sections, and the parameters used are listed in the Table \ref{tab_model_parameters}.  The resulting SED is plotted in the figure \ref{fig_SED_result}.

\subsection{The Inner blob}
\paragraph{Pair-plasma.}
We set the photosphere's velocity as $\beta_{\rm pl, ph} \approx 0.5 \,(\Gamma_{\rm pl} = 1.15)$ and the emission line peaking at $\varepsilon_{\rm pl} \approx 511 \, {\rm keV}$ with a photon distribution given by equation  (\ref{eq_line_distribution}). We take the lower disc luminosity to produce the pair-plasma, i.e, $L_d = 3 \times 10^{-3} L_{E}$ . The photosphere's radius is set at $R_{\rm ph} \approx R_g$. Using these values the corresponding photon energy density is $u_{\rm pl} \approx 3 \times 10^{6} \, (M_{\bullet}/10^9 M_{\odot}) \, {\rm erg \, cm^{-3}}$ in the pair-plasma frame.

\paragraph{VHE gamma-ray emission.}
We require the photopion process to explain TeV gamma-rays, therefore, $\mathcal{D}_i$ must have low value. Here, we choose the value of the boost factor of the inner blob of $\Gamma_i = 3$, with the corresponding the Doppler boost of $\mathcal{D}_i \simeq 6$. Then,  the relative Lorentz factor with the pair-plasma is $\Gamma_{\rm rel} \approx 1.83$ for $\beta_{\rm pl} = 0.5$, and the gamma-rays produced  by photopion are above $\gtrsim 0.3 \, \rm TeV \,$ (see Eq. \ref{eq_TeV_gamma_rays}). Note that a very compact size of this region, i.e., $R_i \sim R_g$, must be associated with a fast observed variability $t_i^{\rm ob} \gtrsim r_g / \mathcal{D}_i \sim  \rm few \;  min$. It is worth noting that these values have been observed in some HBL, e.g., see \cite{Aharonian&etal_2007ApJ...664L..71A,Albert&etal_2007ApJ...669..862A}. However, the variability's timescale in the VHE-band for 1ES 0229 +200 shows indication on yearly timescales \citep{Aliu&etal_2014ApJ...782...13A}. Our model can accommodate the observation only if the dissipation process occur in quasi-stationary during long timescale.

Here, we use a proton distribution that follows a simple power-law function with spectral index of $\alpha_p=1.8$. The lower value predicted by Fermi acceleration was chosen  with the aim to obtain a hard spectrum. Also we assume as the minimum and maximum proton energies as $\epsilon'_{p,\rm min} = 1 \, \rm GeV$ and $\epsilon'_{p,\rm max} = 100 \, \rm TeV$ in the comoving frame, respectively. We do not take a larger proton energy because the photons coming from pair plasma only interact via photopion with protons up to that value (see Figure (\ref{fig_optdepth}a)). Also, this figure shows that secondary electrons produced by photo-pair processes can be neglected compared with the contribution of the photopion processes. The electron cooling is mainly by external inverse Compton (EIC) (see figure \ref{fig_optdepth}a) and a remarkable feature peaking at $\sim$ MeV must emerge by the interaction with the pair-plasma emission. However, the flux is strongly attenuated by the same external pair-plasma radiation (see figure \ref{fig_optdepth}b), therefore we do not consider it too. From figure (\ref{fig_optdepth}b) we can notice that only gamma-rays with energies above than 100 GeV is not internally attenuated.
Therefore, only gamma-rays above $100 \, \rm GeV$ from inner blob must emerge without significant attenuation.


\subsection{The Outer blob}
Here, we assume the outer blob moves with A Lorentz factor of $\Gamma_{\rm o} = 5$ with a Doppler factor of $\mathcal{D}_{\rm o} \simeq 10$. Here we assume that this blob is located between the BLR and DT location. Then radiation are seen in the blob with a redshift and blue-shift effect for BLR and DT, respectively.

\paragraph{Synchrotron radiation.}

The electrons in this region are responsible for the X-ray emission via synchrotron radiation, for 1ES 0229 + 200 which peaks at an energy of $E_{s, \rm pk}^{\rm ob} \sim  10 \, {\rm keV}$ \citep{Costamante2018} and using feasible magnetic field strength in the range of $0.1 - 1 \, \rm G$ we demand energy electron at break be greater than
\begin{equation}
 \gamma_{e,\rm b}^{\prime\prime} \gtrsim 3\times10^5 \, (1+z) \, ( {E_{\rm s}^{\rm ob}}/{{10 \, \rm KeV}} )^{1/2}  \left({\mathcal{D}_o}/{10}\right)^{-1/2} \left({B_o''}/{0.5 \, {\rm G}}\right)^{-1/2}\, .
\end{equation}

If the size of the outer region is associated with a daily variability, then we have $R_{\rm o} \approx 2.6 \times 10^{16} \, {\rm cm} \, (t_{\rm var}/{\rm day})(\mathcal{D}_{\rm o}/10)$. Furthermore, the energy density produced by synchrotron process is
\begin{align}\label{eq_syn_density}
    u_s^{\prime\prime} 
    &\sim 
    10^{-2} \, {\rm {erg}\, {cm^{-3}}} \, ( {\mathcal{D}_o}/{10} )^{-4} ( {R_{\rm o}^{\prime\prime} }/{10^{16} \, \rm cm} )^{-2} \, ( {L_s^{\rm ob} }/{10^{45} \, \rm erg \, s^{-1}} ) \,.
\end{align}

\paragraph{The BLR and DT radiation field.}

We set the parameter $\phi_{\rm BL} = \phi_{\rm DT} = 0.1 $, then we have a BLR luminosity of $L_{\rm BL} \approx 3 \times 10^{\rm -4} L_E$.  This value lies in the range of BL Lacs proposed by \citep{Ghisellini&Tavecchio_2008}. Note that this value is similar to that reported in other EHBLs with possible high-energy neutrino association $<3 \times 10^{-4} \, L_E$ \citep{Giommi&etal_2020}. From equation (\ref{eq_rBL}) and (\ref{eq_energy_densities}), the BLR's location and DT's radius are $r_{\rm BL} \approx 7 \times 10^{16} \, {\rm cm} \, ( M_{\bullet}/10^9 M_{\odot} )^{0.5}$ and $r_{\rm DT} \approx 10^{18} \, {\rm cm} \, ( M_{\bullet}/10^9 M_{\odot} )^{0.5}$, respectively. Its energy densities are
\begin{equation}\label{eq_energy_density_in_outer_blob}
    u''_{\rm BL} \approx 1.9 \times 10^{-4} {\rm erg}{\, \rm cm^{-3}}\, ( {\mathcal{D}_o}/{10} ) ^{-2}  \, ({\phi_{\rm BL}}/{0.1)} \, ,
    \quad
    u''_{\rm DT} \approx 4.3 \times 10^{-2} \, {\rm erg}{\rm \, cm^{-3}}\, ( {\mathcal{D}_o}/{10}) ^{2}  \, ({\phi_{\rm BL}}/{0.1}) \,.
\end{equation}
Finally, the peak energies are $\varepsilon''_{\rm BL} \approx 1 \, (\mathcal{D}_o/10)^{-1} \, \rm eV$ and $\varepsilon''_{\rm DT} \approx 1 \, (\mathcal{D}_o/10) \, \rm eV$ for BLR and DT, respectively. 

.

\begin{table}
\caption{Parameters used in our model for the EHBL/TBL 1ES 0229 +200. All quantities are expressed in the respective comoving frame.}
\begin{tabular}{c c c c  | c c c c c c c}
\hline
   \multicolumn{4}{c}{\textbf{ Inner Blob }} &  \multicolumn{6}{c}{\textbf{ Outer Blob }}
\\
\hline
$B$    & $\mathcal{D}$     & $R$  & & $B$    & $\mathcal{D}$     & \multicolumn{2}{c}{$R$}
\\
$20 \; \rm G$    & $6$     & $2.3 \times 10^{14} \; \rm cm$  & & $0.19 \; \rm G$    & $10$     & \multicolumn{2}{c}{$1.3 \times 10^{16}  \; \rm cm$}
\\ &&&&
\\
$K_p$ & $E_{\rm p,min}$  & $E_{\rm p,max}$ & $\alpha$ &  $K_e$ & $\gamma_{\rm e,min}$  & $\gamma_{\rm e,b}$ & $\gamma_{\rm e,max}$ & $\alpha_1$ & $\alpha_2$
\\
$2e40$ & $ 1 \; {\rm GeV}$  & $100 \; {\rm TeV}$ & $1.8$ & $9e50$ & $1$  & $3e5$ & $5e6$ & $1.8$ & $3.1$
\\&&&&
\\
\multicolumn{2}{c}{$L_p$} & \multicolumn{2}{c|}{$L_B$} &  \multicolumn{2}{c}{$L_e$} & \multicolumn{2}{c}{$L_B$} &
\multicolumn{2}{c}{$L_B / L_p$} 
\\
\multicolumn{2}{c}{$9\times 10^{45} \; \rm erg \, s^{-1}$} & \multicolumn{2}{c|}{$6\times 10^{42} \; \rm erg \, s^{-1}$} &  \multicolumn{2}{c}{$5.6\times 10^{43} \; \rm erg \, s^{-1}$} & \multicolumn{2}{c}{$1.4\times 10^{43} \; \rm erg \, s^{-1}$} & 
\multicolumn{2}{c}{$0.25$} & 
\\
\hline
\end{tabular}\label{tab_model_parameters}
\end{table}

\begin{figure}
\begin{minipage}[b]{\linewidth}
\centering
\includegraphics[width=0.9\linewidth]{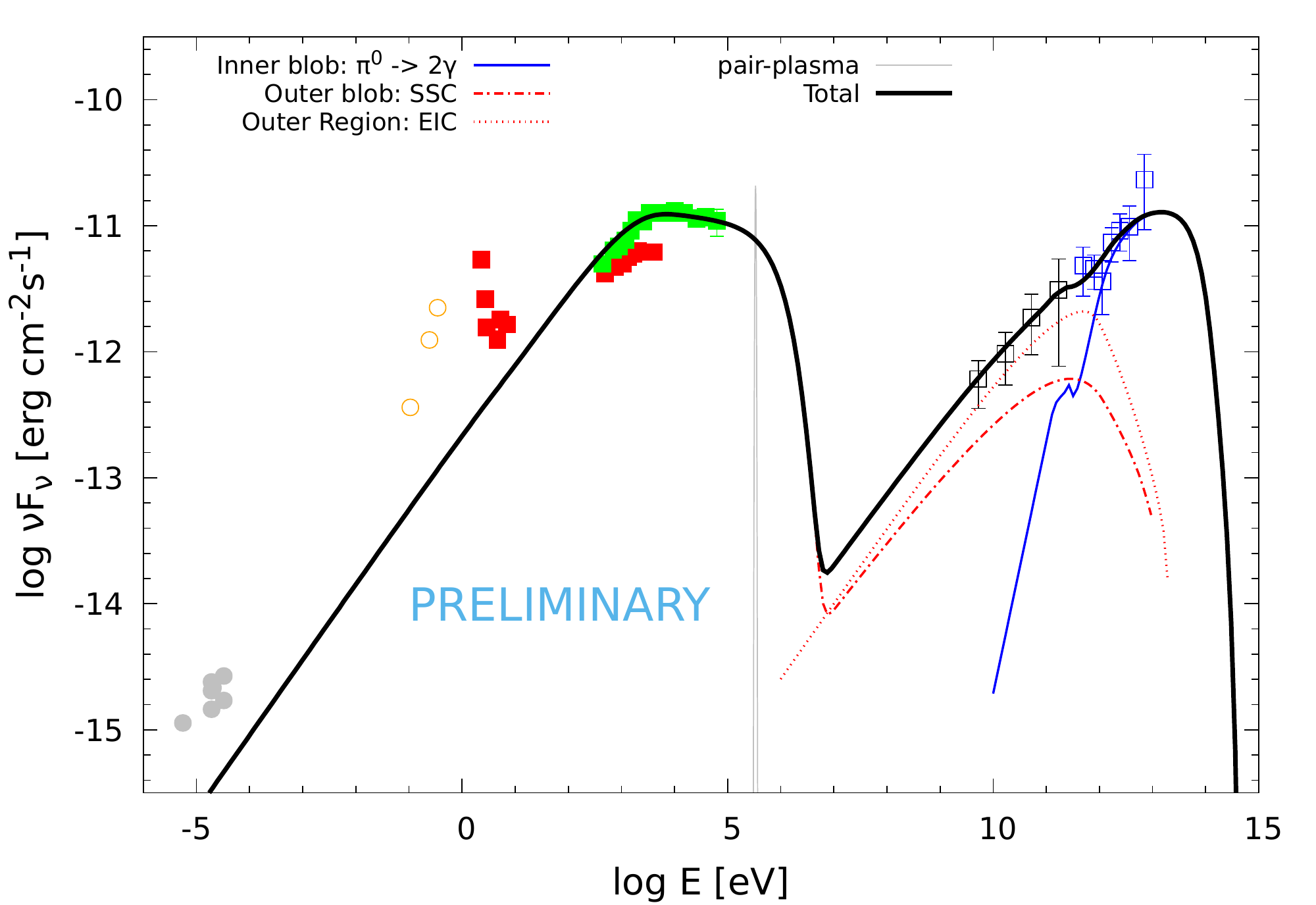}
\end{minipage}\hfill 
\caption{The broadband spectral energy distribution  of 1ES0229+200 described with our multizone model. The outer blob is modeled using the SSC and EIC processes (blue lines) and the inner blob using the photopion process (blue line).}
\end{figure}\label{fig_SED_result}

\paragraph{High-Energy emission.}
From equations (\ref{eq_syn_density}) and (\ref{eq_energy_density_in_outer_blob}) we show that the BLR radiation can be discarded for Compton scattering in the outer blob. Notice that EIC scattering could be dominated by DT respect to synchrotron by a factor as high as $L^{\rm DT}_{ic} / L_{ssc} \sim 4$.  All electrons with $\gamma_e^{\prime\prime}  \ll 5\times 10^5 ( \epsilon_{\rm DT}/ 0.1 {\rm eV} )^{-1} (\mathcal{D}_o/10)^{-1} $ scatter off in Thompson regime with the dusty torus radiation, and the Klein-Nishina (KN) suppression is avoided. For instance, if we choose $\gamma_{e,\rm br}^{\prime\prime}  = 3\times 10^5$ the peak of EIC is expected around $E_{ic}^{\rm ob} \sim 0.48 \, {\rm TeV} (\gamma_e^{\prime\prime  } / 2 \times 10^5)^2 (\mathcal{D}_o/10)^2 (\epsilon_{\rm DT}/0.1 \, {\rm eV})$ just below the gamma-ray peak produced by photopion decay products in the inner blob. 

\section{Discussion and Conclusion}\label{sec:Con_&_Dis}
The extreme BL Lac behaviour possesses challenges for modelling them with either a pure leptonic or hadronic model with one-zone emission because they demand extreme parameter values. In this work, we have proposed a lepto-hadronic model with two-zone emission regions and external radiation sources in order to explain the broadband spectral energy distribution for the prototype extreme blazar 1ES 0229 +200.  We showed that external radiation from a pair-plasma and dusty torus plays an important role to produce a hard spectrum at very-high energies.


The very-high energy component is produced mainly by two-contributions: an hadronic and a leptonic: The hadronic is produced by the photopion process of pair-plasma's photons with accelerated protons inside the inner blob producing the highest energy gamma-rays above TeV-energies. The leptonic is produced by SSC model and by EIC with the dusty torus's photons inside the outer blob. Finally, the SED's low-energy bump is produced only by synchrotron radiation of the electron population in the outer blob. 

Our model provides a good fit without extreme parameter values as other models require (see Table \ref{tab_model_parameters} and figure (\ref{fig_SED_result}). Our model, demands a minimum electron's energy with a low value and provides a  magnetization's value close to the unity in the outer blob and in the inner blob the proton luminosity required is a sub-Eddington limit. Finally, it is worth noting that if the pair-plasma or an inner blob are not there, the TBL behaviour disappears conserving only a EHBL one. This agrees with the existence of blazars that are EHBL but no TBL or viceversa.

\acknowledgments
This work was made with the support of DGAPA-UNAM thanks to the grants IG101320 and IN106521.

\bibliographystyle{JHEP} 
\bibliography{biblio}

\end{document}